\documentclass[journal=apchd5,manuscript=article,layout=traditional]{achemso}

\usepackage[version=3]{mhchem} 

\title{Optical properties of Ag, Au, and Cu from first principles}

\author{Xiao Zhang}
\affiliation{Department of Materials Science and Engineering, University of Michigan, Ann Arbor, MI, 48109, USA}
\author{Emmanouil Kioupakis}
\affiliation{Department of Materials Science and Engineering, University of Michigan, Ann Arbor, MI, 48109, USA}
\email{kioup@umich.edu}

\begin{document}






\begin{abstract}
We present a comprehensive framework for investigating the optical response of metals from first principles that combines density functional theory, many-body perturbation theory, and efficient interpolation techniques based on maximally localized Wannier functions, and apply it to analyze the optical properties of silver (Ag), gold (Au), and copper (Cu). We evaluate the optical properties of these metallic materials considering both single-particle direct and phonon-assisted excitations, as well as the resistive Drude contribution. We find an overall excellent agreement with experimental optical measurements for these materials, and show that both single-particle and collective excitations are important in capturing their optical response in the infrared. Our methodology provides fundamental understanding of the optical response of metals and is generally applicable to investigate the optoelectronic properties of emerging metallic materials. 
\end{abstract}

\section{Introduction}

Understanding optical properties of metallic systems are crucial in understanding their applicability in applications. For example, the high conductivity of metals result in strong plasmon oscillations due to the negative refractive index\cite{west_searching_2010}, rendering applications of noble metals such as Ag and Au in sensing, waveguide, solar cells, and bio-imaging.\cite{novotny_principles_2012,kolesov_waveparticle_2009,ozbay_plasmonics_2006,maier_plasmonicsroute_2001,maier_plasmonics_2005,huang_cancer_2006,schuller_plasmonics_2010,ferry_design_2010} Gold nanoparticles, for example, have been widely utilized in applications including medical imaging, diagnosis, treatment\cite{huang_cancer_2006,kah_early_nodate,sharifi_plasmonic_2019}, toxic detection\cite{priyadarshini_gold_2017,cho_ultrasensitive_2012}, and enhanced Raman spectroscopy. \cite{schlucker_surface-enhanced_2014}

Despite the success of noble metals in applications, consistent first-principles framework for understanding the optical properties of metallic systems is still needed. 
Modeling optical responses in metals, especially in the infrared region, is challenging due to the presence of free carriers. 
While techniques for computational characterization of optical response from first principles has been well-established for semiconductors, free carriers play a crucial role in the optical absorption of metals, especially in the infrared region. 
Previous studies have examined free-carrier absorption in doped semiconducting systems including silicon\cite{zhang_ab_2022}, nitrides\cite{peelaers_free-carrier_2015}, and transparent conducting materials\cite{peelaers_phonon-_2019,peelaers_fundamental_2012}. 
The significant difference in electrical conductivity of doped semiconductors and metallic systems renders it valuable to investigate absorption in the metallic systems with considering both single-particle excitations as well as the semi-classical resistive contribution. 
Single-particle contributions arise from the excitation of electrons from occupied to empty states, which can occur via direct transitions (absorption of a photon) or indirect transitions (photon absorption and electron scattering with a phonon). 
In contrast, resistive contributions result from collective oscillations of electrons near the Fermi level. 
Many studies have focused on understanding the optical properties in metallic systems, either considering direct absorption or intraband contributions within the semi-classical Drude model,\cite{maksimov_first-principles_1988,li_first-principles_2016,kandemir_optical_2024,peng_electronic_2016,prandini_photorealistic_2019,stahrenberg_optical_2001,marini_dynamical_2003}, or focusing on a different perspective such as hot-carrier dynamics\cite{brown_nonradiative_2016,PRXEnergy.1.013006,10.1021/acs.jpcc.3c05347,biswas2024exciton,richter2020ultrafast}. 

In this work, we present a comprehensive framework to study of the optical response of metallic materials, and apply it to compare to experimental data for the elemental noble metals gold (Au), silver (Ag), and copper (Cu) considering both single-particle and resistive contributions.
Using density functional theory\cite{hohenberg_inhomogeneous_1964,kohn_self-consistent_1965} (DFT) and many-body perturbation theory (MBPT), we achieve excellent agreement with experimentally measured electronic structure of these metallic systems. 
Using density functional perturbation theory\cite{baroni_phonons_2001} (DFPT) we accurately characterize phonon dispersion relationship of the materials. We compute electrical conductivity using the maximally localized Wannier functions technique\cite{marzari_maximally_2012} to interpolate to fine Brillouin zone sampling grids, and iteratively solving the Boltzmann transport equation. 
We then evaluate the optical absorption, with the direct and phonon-assisted contribution obtained through first- and second-order time-dependent perturbation theory, and the resistive contribution obtained using the calculated electrical conductivity. We show that this framework results in accurate quantitative description of the optical response of metals across a broad range of spectral regions. 

\section{Computational methods}

First-principles calculations for ground-state electronic properties are performed using density functional theory (DFT)\cite{hohenberg_inhomogeneous_1964,kohn_self-consistent_1965}. 
We used experimentally measured lattice constants for all materials considered: Ag: 4.08 \r{A}\cite{davey_precision_1925}, Au: 4.079\r{A}\cite{maeland_lattice_1964}, and Cu: 3.61 \r{A}\cite{straumanis_lattice_1969}. 
The ground-state calculations are performed with the Quantum Espresso\cite{giannozzi_quantum_2009,giannozzi_advanced_2017} (QE) package. The PBEsol\cite{perdew_restoring_2008} exchange-correlation functional is used following Ref. \citenum{sundararaman_theoretical_2014}, and the optimized Vanderbilt norm-conserving pseudopotentials are obtained from PseudoDojo\cite{hamann_optimized_2013,van_setten_pseudodojo_2018}. 
The wave functions are expanded into plane waves with a cutoff energy of 100 Ry. 
Ground-state calculations are performed with a Brillouin zone (BZ) sampling grid of $14\times14\times14$, displaced by half of a grid spacing to achieve convergence of the total energy within 1 meV/atom. 
The phonon properties are calculated with DFPT\cite{baroni_phonons_2001} using QE\cite{giannozzi_quantum_2009,giannozzi_advanced_2017}. 
For phonon properties, PBEsol exchange-correlation functional is used for Ag and Au, and the LDA exchange-correlation functional is used for Cu. 
The choice of LDA exchange-correlation functional for Cu for only phonon properties is based on better agreement of the phonon dispersion relationship compared to experimental measurements as illustrated in Figure~\ref{fig:phonon} later in this paper. The electron-phonon matrix elements are calculated with a BZ sampling grid of $6\times6\times6$.

For band-structure calculations, we include quasiparticle corrections with the GW approximation. 
The calculations are performed with the BerkeleyGW\cite{deslippe_berkeleygw_2012,deslippe_coulomb-hole_2013} package. The random phase approximation, extended to finite frequencies using the generalized plasmon pole model from Hybertsen and Louie\cite{hybertsen_electron_1986} is used to evaluate the dielectric screening. The static remainder approach is used to reduce the number of empty orbitals required.\cite{deslippe_coulomb-hole_2013}
A cutoff of 40 Ry is used for evaluating the screened Coulomb interaction. 
The GW calculations are performed with BZ-sampling grid of $12\times12\times12$ for all materials. 
We used the PBEsol+$U$ corrections\cite{liechtenstein_density-functional_1995} as a starting point to obtain the correct position of the metal d-states. The $U$ values used are 1 eV for Au and Ag, and 2 eV for Cu. The values are chosen based on agreement between the electronic structure from angle-resolved photoemission experiments\cite{wehner_valence-band_1979,mills_angle-resolved_1980,knapp_experimental_1979,courths_angle-resolved_1983} and our calculated electronic band structure after quasiparticle corrections are taken into account. 
We show the justification of the choice of the $U$ values in supplemental information Figure S1.

The quasiparticle energies, electron-phonon matrix elements, and velocity matrix elements are subsequently interpolated to fine BZ sampling grids, detailed in next paragraph, via the maximally localized Wannier functions technique. 
To evaluate the electrical conductivity, we solve the Boltzmann transport equation iteratively\cite{marzari_maximally_2012,ponce_first-principles_2020,ponce_towards_2018,lee_electronphonon_2023}, and test for convergence of the conductivity with respect to the BZ samplings. 

Optical properties from single-particle contributions are evaluated with second-order perturbation theory\cite{zhang_ab_2022,noffsinger_phonon-assisted_2012}, with details of the inclusion of free-carrier contributions reported in Ref. \citenum{zhang_ab_2022}. 
The electron-photon and the electron-phonon interactions are considered as perturbations, and we obtain transition rates using time-dependent perturbation theory. 
A divergence is present in evaluating phonon-assisted optical properties if direct and indirect transition are resonant with each other\cite{zhang_ab_2022,brown_nonradiative_2016}, and the divergence is compensated via the inclusion of a numerical broadening parameter. 
We choose a broadening parameter of 0.1 eV, and demonstrate that the phonon-assisted spectra are insensitive to the value of the broadening parameter in the spectral region where phonon-assisted processes dominate (See supplemental information Figure S2 for an illustration). 
To evaluate phonon-assisted optical absorption, a BZ sampling grid of $48\times48\times48$ is used. To evaluate direct and resistive contribution, BZ sampling grids of $100\times100\times100$ is used. For Au, the grid is increased to $180\times180\times180$ to ensure a smooth spectrum.  

\section{Results and discussion}

\subsection{Electronic structure and phonon dispersion}

\begin{figure*}
    \centering
    \includegraphics[width=\textwidth]{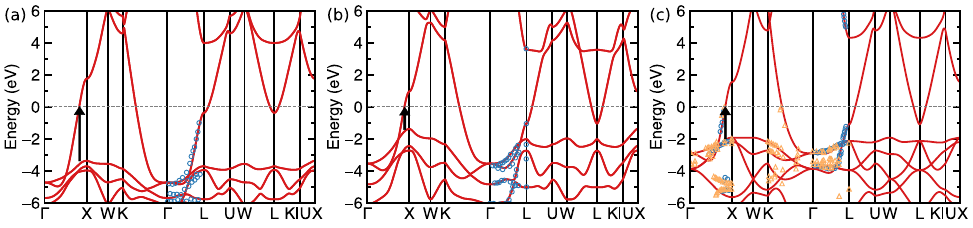}
    \caption{Electronic band structure, calculated with PBEsol+U+GW, of (a) Ag, (b) Au, and (c) Cu. Spin-orbit coupling is included for Ag and Au. The black arrows mark the lowest direct transition for Ag (3.69 eV), Au (1.63 eV), and Cu (2.09 eV). 
    The calculated electronic band structures (red solid lines) agree well with experimental measurements, illustrated with marks: (a) Ref.\citenum{wehner_valence-band_1979} (b) Ref.\citenum{mills_angle-resolved_1980} (c) Circles: Ref.\citenum{knapp_experimental_1979}, Upper triangles marks: Ref.\citenum{courths_angle-resolved_1983}.
    }
    \label{fig:bands}
\end{figure*}

Figure~\ref{fig:bands} shows the calculated electronic band structures of Au, Ag, and Cu evaluated with the PBEsol+$U$+$GW$ approach. 
The calculated bands agree well with experimental measurements\cite{wehner_valence-band_1979,mills_angle-resolved_1980,knapp_experimental_1979,courths_angle-resolved_1983}, particularly near the Fermi level, which is crucial for accurately evaluating the electrical conductivity. 
In addition, the combination of PBEsol, Hubbard $U$ corrections and the $GW$ approximation ensure good agreement of both the occupied and the empty states with experimentally measured electron energies, which is essential to correctly evaluate the onset of direct optical absorption. 
The minimum direct transition for all three metals occurs along the $\Gamma\rightarrow X$ direction, which is also in good agreement with experimental measurements from ARPES. 

We next investigate the effects of spin-orbit coupling (SOC) for Ag and Au. 
A comparison between the band structures with and without SOC is shown in the supplemental information Figure S3. 
The SOC effect is stronger in Au, due to its higher atomic number. 
At the $L$-point, the band splitting due to SOC at the VBM is 0.23 eV for Ag and 0.75 eV for Au, and at the $X$-point, the splitting is 0.34 eV for Ag and 1.07 eV for Au. 
These splittings induced by SOC for Au and Ag are necessary for obtaining good agreement with experimental measurements up to 5 eV below the Fermi level. 
Therefore, the inclusion of SOC is essential to correctly capture optical absorption in the infrared and close to the optical onset of direct absorption. 
SOC has an insignificant effect for the s-orbital bands near the Fermi level for both Au and Ag, and hence was not considered for the evaluation of electrical conductivity due to the higher computational cost. 
For Cu, SOC causes a splitting is only around 0.1 eV at the VBM (shown in Figure S3), therefore is not included in our analysis as this splitting is comparable to the broadening parameters in e.g. transport and optical spectra calculations. 

\begin{figure*}
    \centering
\includegraphics[width=\textwidth]{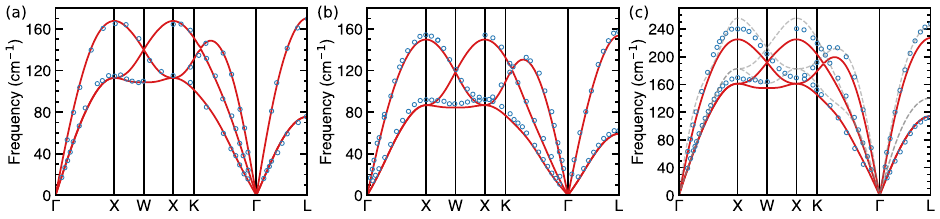}
    \caption{Phonon dispersion relationship calculated with Quantum Espresso using DFPT for (a) Ag, (PBEsol) (b) Au, (PBEsol) and (c) Cu. (red: LDA, grey: PBEsol) Our calculated phonon dispersions (red solid lines) agree well with experimental measurements (open circles) from (a) Ref.\citenum{kamitakahara_crystal_1969}, (b) Ref.\citenum{lynn_lattice_1973}, (c)Ref.\cite{svensson_crystal_1967}. For Cu, a larger discrepancy in the dispersion is observed when PBEsol is used, therefore LDA is used for phonon-related properties in subsequent calculations.}
    \label{fig:phonon}
\end{figure*}
The phonon dispersion relationships of all three materials are shown in Figure~\ref{fig:phonon}. 
The dispersion curves exhibit similar trends across the Brillouin zone, with Au demonstrating the lowest phonon frequencies and Cu the highest, consistent with their respective atomic masses. 
Our calculated frequencies closely align with experimental measurements, showing maximum deviations of 1.2\%, 2.8\%, and 6.2\% for Ag, Au, and Cu, respectively.

\subsection{Electrical conductivity}

\begin{figure}
    \centering
    \includegraphics[width=0.8\columnwidth]{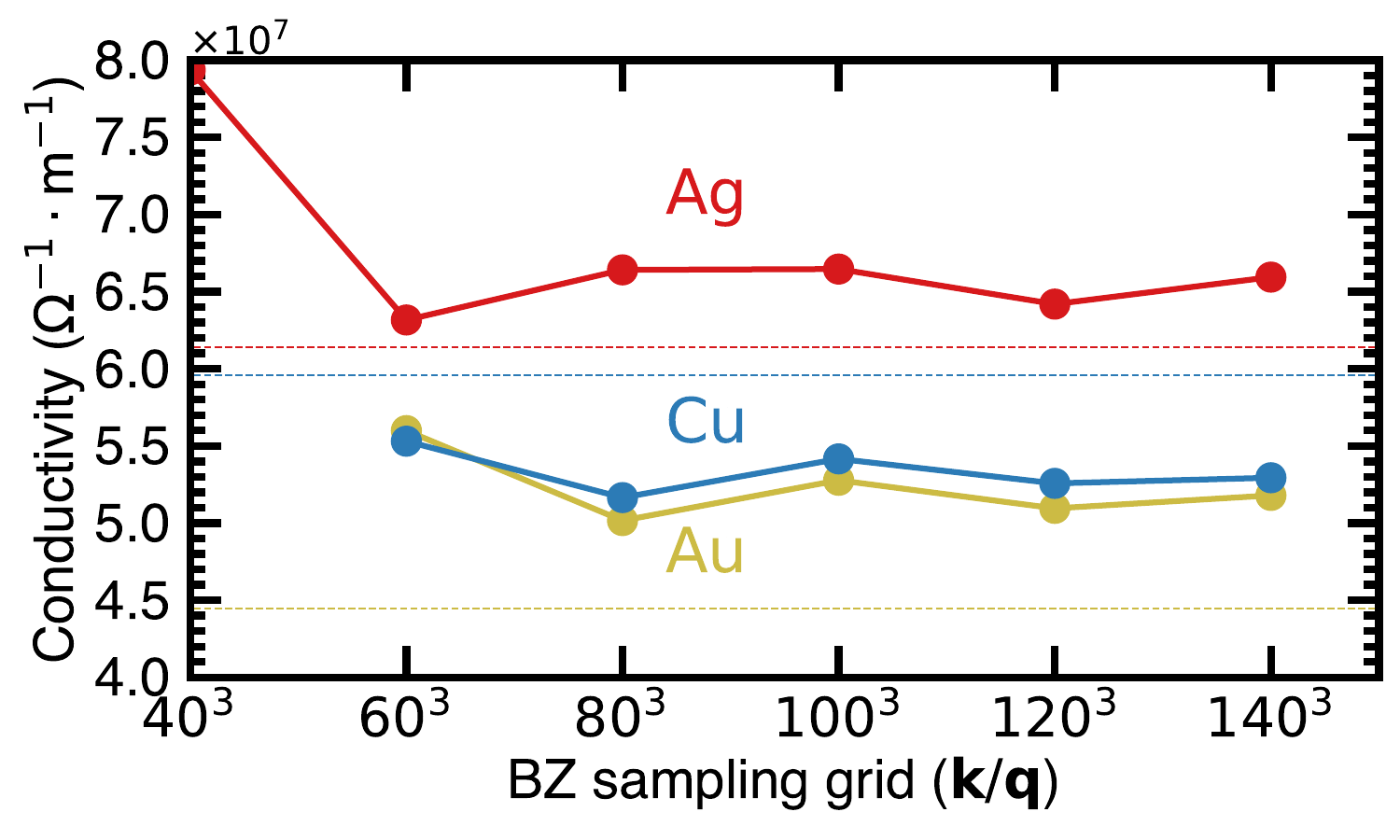}
    \caption{Convergence of electrical conductivity calculated with Wannier interpolation for Ag (red), Au (olive), and Cu (blue). Experimentally measured values are shown as horizontal dashed lines, taken from the recommended values of Ref.\citenum{matula_electrical_1979} for Ag, Au, and Cu. For all three materials, the calculated conductivity agrees with experimental measurements within 10\% for the most converged sampling.}
    \label{fig:mobility}
\end{figure}

We evaluate the electrical conductivity of all three materials by iteratively solving the Boltzmann transport equation. 
The convergence of the electrical conductivity versus the BZ sampling grid is shown in Figure~\ref{fig:mobility}. 
For all three metals, the calculated electrical conductivity converges within 10\% for a grid size of 100×100×100. 
Our converged simulated results also agree within 10\% comparing to experimentally measured values in Ref. \citenum{matula_electrical_1979}. 
As illustrated from Ref. \citenum{matula_electrical_1979}, a $\sim$10\% variation is also seen in the experimentally measured values. 
Our previous work\cite{zhang_ab_2022} has shown that such a variation in electrical conductivity does not strongly affect the evaluation of the resistive contribution to the optical response.  

\subsection{Optical properties}

\begin{figure}
    \centering
    \includegraphics[width=0.8\columnwidth]{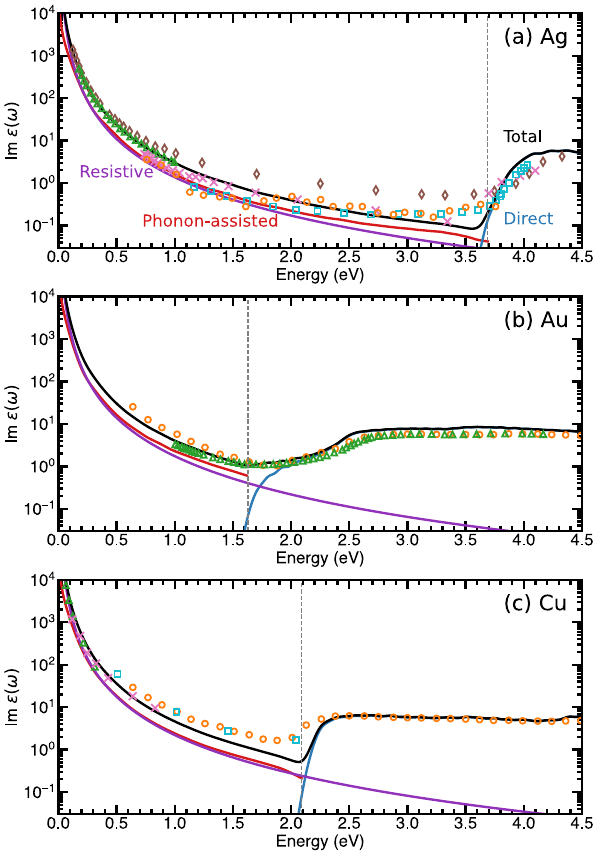}
    \caption{Imaginary part of the dielectric function from phonon-assisted contribution (red), direct contribution (blue), resistive contribution (purple), and total contribution (black) of (a) Ag, (b) Au, and (c) Cu. Experimental measurements are shown as marks from (a) Ref.\citenum{johnson_optical_1972} (orange circles), Ref.\citenum{choi_optical_2020} (green triangles), Ref.\citenum{mcpeak_plasmonic_2015} (pink crosses), Ref.\citenum{wu_intrinsic_2014} (cyan squares), and Ref.\citenum{rakic_optical_1998} (brown diamonds). (b) Ref.\citenum{johnson_optical_1972} (orange circles) and Ref.\citenum{olmon_optical_2012} (green triangles). (c) Ref.\citenum{johnson_optical_1972} (orange circles), Ref.\citenum{lenham_applicability_1966} (green triangles), Ref. \citenum{dold_optische_1965} (pink crosses), and Ref.\citenum{hagemann_optical_1975} (cyan squares). Excellent agreement can be seen for all three materials.  }
    \label{fig:optics_imepsilon}
\end{figure}

We evaluate the imaginary part of the dielectric function considering direct absorption, phonon-assisted absorption, and resistive contribution. Figure~\ref{fig:optics_imepsilon} shows the calculated imaginary part of the dielectric function [Im $\varepsilon(\omega)$] for Ag, Au, and Cu in the range of infrared to visible light. 
The phonon-assisted contribution is not considered for photon energies beyond the direct-absorption onset, which is 3.69 eV for Ag, 1.63 eV for Au, and 2.09 eV for Cu. 
This is due to the fact that direct transitions dominate in the spectral range where they are allowed, while the phonon-assisted spectra are affected by the artificial sensitivity of second-order perturbation theory to the broadening parameter. 
For photon frequencies below the calculated direct-absorption onset, the phonon-assisted contribution, which is a single-particle excitation mechanism, and the resistive contribution, which results from the collective oscillation of the free electrons, are comparable in magnitude. 
Therefore, both contributions are required to correctly evaluate the optical response. 
On the other hand, for photon frequencies beyond the direct onset, absorption occurs primarily via interband absorption from the occupied to the empty orbitals, while the other contributions are negligible. 
The total Im $\varepsilon(\omega)$ considering both single-particle excitations and resistive contributions is in excellent agreement compared to experimental measurements for all three metals. 
For silver, the experimentally measured data exhibit a higher variance, and our results agree particularly well with more recent reports such as Ref.\citenum{choi_optical_2020,mcpeak_plasmonic_2015,wu_intrinsic_2014}, in which high-quality single-crystalline films were shown to reduce loss and lower the dielectric function in the IR region. 
This agreement confirms that none of the contributions can be neglected when evaluating optical absorption over the full photon-energy range. 

The calculated spectra also reveal distinct material-dependent features. Specifically, silver and copper display relatively sharp direct absorption onsets corresponding to interband transitions near the $L$ and $X$ points, whereas gold shows a much broader onset. 
This broadening originates from strong spin–orbit coupling in the Au $5d$ states, which lifts degeneracies near the top of the occupied bands and smears the direct onset over a wide energy range. 
As shown in our supplemental information Figure S4 and S5, including spin–orbit coupling is critical to quantitatively predict this behavior in gold and to correctly describe the experimentally measured spectra in the NIR–visible range. 

\begin{figure}
    \centering
    \includegraphics[width=0.8\columnwidth]{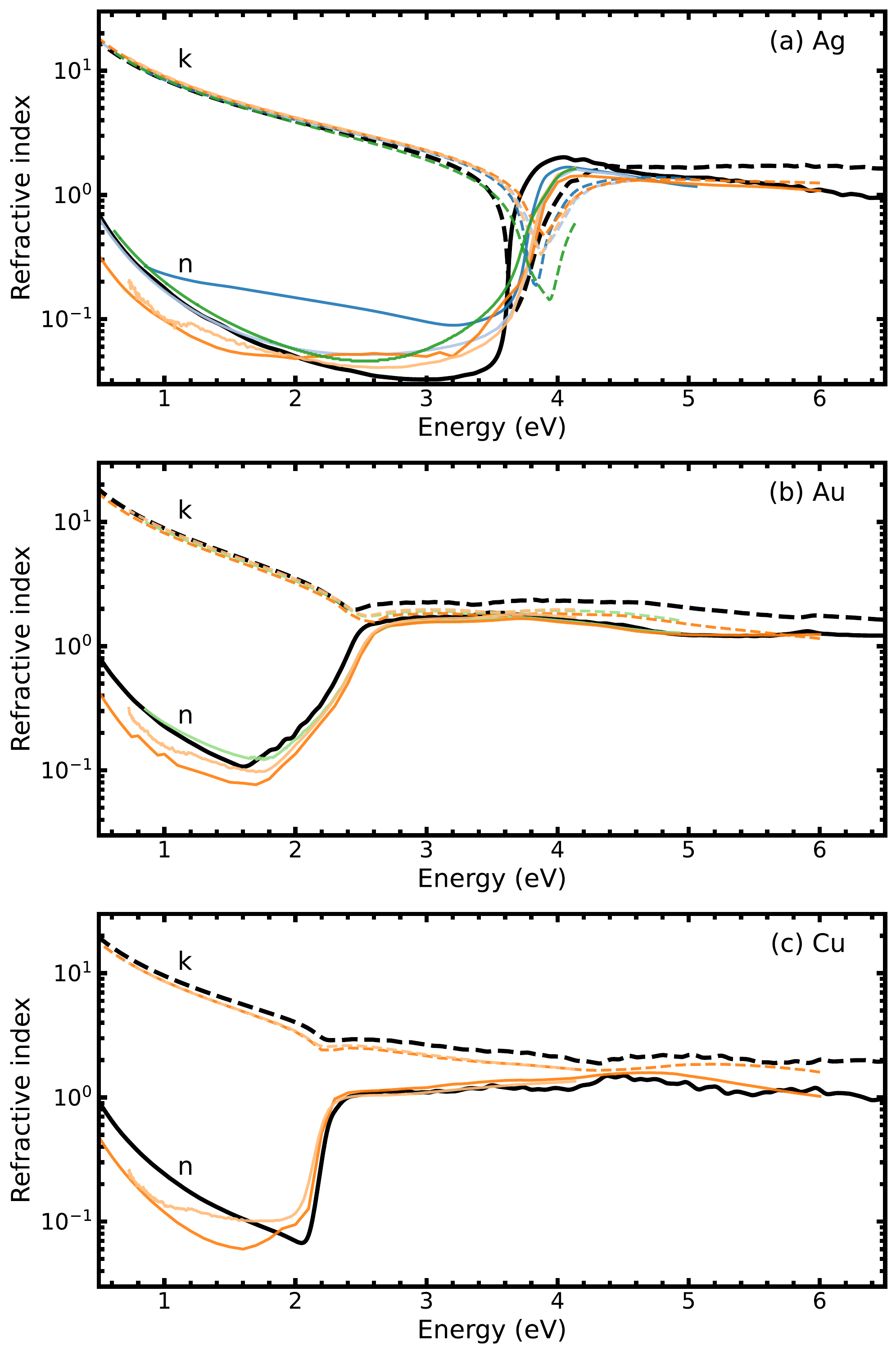}
    \caption{Complex refractive index (black curves) evaluated for (a) Ag, (b) Au, and (c) Cu using the total dielectric function including both the single-particle contribution and the resistive contribution. Excellent agreement for both the shape and the magnitude of the spectra is observed compared to experimental measurements in Ref. \citenum{ferrera_temperature-dependent_2019} (blue), Ref. \citenum{jiang_realistic_2016} (green), Ref. \citenum{yang_optical_2015} (light blue), Ref. \citenum{mcpeak_plasmonic_2015} (light orange), and Ref.\citenum{babar_optical_2015} (orange). 
    }
    \label{fig:optics_refr}
\end{figure}

We further evaluate the complex refractive index of the three metals. 
The real part of the dielectric function is obtained from the imaginary part via the Kramers-Kronig relationship, and the complex refractive index is evaluated via the relationship $[n(\omega)+ik(\omega)]^2=\text{Re }\varepsilon(\omega)+i\text{Im } \varepsilon(\omega)$. 
We show the calculated complex refractive index for all three materials in Figure~\ref{fig:optics_refr}. 
Compared to recently published experimental measurements from the past decade\citenum{mcpeak_plasmonic_2015,ferrera_temperature-dependent_2019,jiang_realistic_2016,yang_optical_2015,babar_optical_2015}, our calculated results show excellent agreement. 
The excellent agreement in the spectral regime below the direct onset further illustrates the importance of including both single-particle and resistive contributions. 
In the region close to the direct absorption onset, the kinks in the spectra are well-predicted as the main contribution for optical absorption changes from phonon-assisted and resistive to direct absorption. 
In the spectral region above the direct onset, direct absorption dominates over the other contributions. 
The correct description of the electronic band structure resulted from the PBEsol+$GW$+$U$ approach, and the inclusion of spin-orbit coupling for Ag and Au play an important role in the correct description of optical absorption near and beyond the direct absorption onset. 

\begin{figure}
    \centering
    \includegraphics[width=0.8\columnwidth]{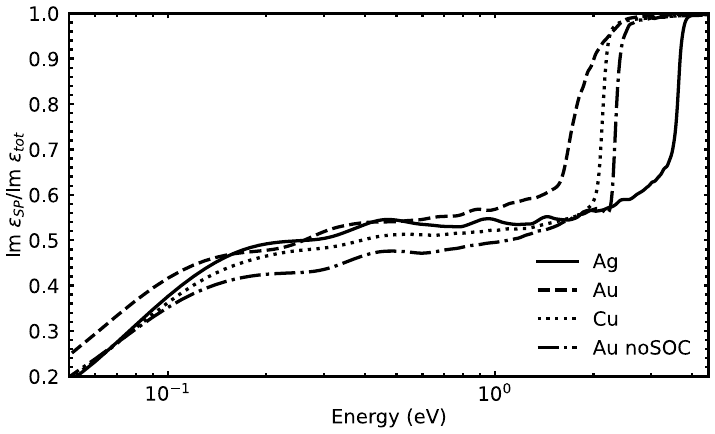}
    \caption{Ratio of single-particle absorption contribution to the total absorption characterized by the imaginary part of the dielectric function for Ag, Au, and Cu. The fraction of single-particle absorption drops significantly for photon energies below 0.4 eV and remains on the order of 50\% in the region of 0.4 eV until the direct onset. The ratio for Au is strongly affected by the inclusion of spin-orbit interactions. }
    \label{fig:ratio}
\end{figure}

The separation of different mechanisms provides insight into applications in plasmonics. 
Previous studies have demonstrated potential enhancement in hot-carrier generation due to the interplay between single-particle excitations and plasmonic oscillations\cite{ma_interplay_2015}. 
Understanding the contribution from single-particle excitation thus provides valuable information on applications such as photocatalysis, where hot-carrier generation is essential.\cite{ma_interplay_2015,vanzan_theoretical_2024,gieseking_review_2016}
In Figure~\ref{fig:ratio}, we plot the ratio of the single-particle contribution to the total imaginary part of the dielectric function for the three noble metal materials as a function of photon energies. 
Below 0.4 eV, the ratio drops significantly due to the strong resistive contribution, while above this energy, single-particle contributions gradually increase. 
In the region above 0.4 eV, Au surpasses the other two materials, exhibiting the highest ratio of single-particle contributions. For the spectral region close to the direct absorption onset, the ratio of single-particle contributions significantly increases as first-order direct absorption is enabled. 
The effect of SOC is particularly evident in gold, for which the onset of strong single-particle contributions shifts from $\sim$2.2 eV to $\sim$1.6 eV with the inclusion of SOC. 
While a comprehensive assessment for practical photocatalysis applications requires considering specific nanoparticle geometries, our method provides a reliable theoretical tool to investigate the different mechanisms of optical absorption in plasmonic materials from a fundamental perspective.  

\section{Conclusion}

In this work, we report a computational framework of characterizing the optical response of metallic systems from first principles, considering both single-particle direct and indirect processes, as well as the semi-classical resistive contribution. We adopt an approach based on second-order perturbation theory to evaluate the indirect phonon-assisted contribution. We show that our approach provides an accurate description for optical response of noble metals Ag, Au, and Cu, in good agreement with experimental optical measurements. Our study suggests the importance of both single-particle absorption and resistive loss in the optical response of metallic systems. Further, our methodology allows a detailed analysis of the various mechanisms for optical absorption in metallic systems at different photon energies. Our results provide benchmark optical spectra for noble metals and a computational tool for accurately modeling dissipation mechanisms in metals, offering a foundation for future studies of emerging complex metallic systems for applications in fields such as plasmonics and photocatalysis. 

\begin{acknowledgement}
This work is supported as part of the Computational Materials Sciences Program funded by the U.S. Department of Energy, Office of Science, Basic Energy Sciences, under Award No. DE-SC0020129. The ground state properties and quasiparticle electronic structures, used resources of the Argonne Leadership Computing Facility, which is a DOE Office of Science User Facility supported under Contract DE-AC02-06CH11357. Computational resources for conductivity and optics calculations were provided by the National Energy Research Scientific Computing Center, which is supported by the Office of Science of the U.S. Department of Energy under Contract No. DE-AC02-05CH11231. 
\end{acknowledgement}
\clearpage

\renewcommand{\thefigure}{S\arabic{figure}}
\setcounter{figure}{0}

\section{Supplemental Information}

\subsection{Effects of $U$ parameter}

\begin{figure}[!ht]
    \centering
    \includegraphics[width=0.7\columnwidth]{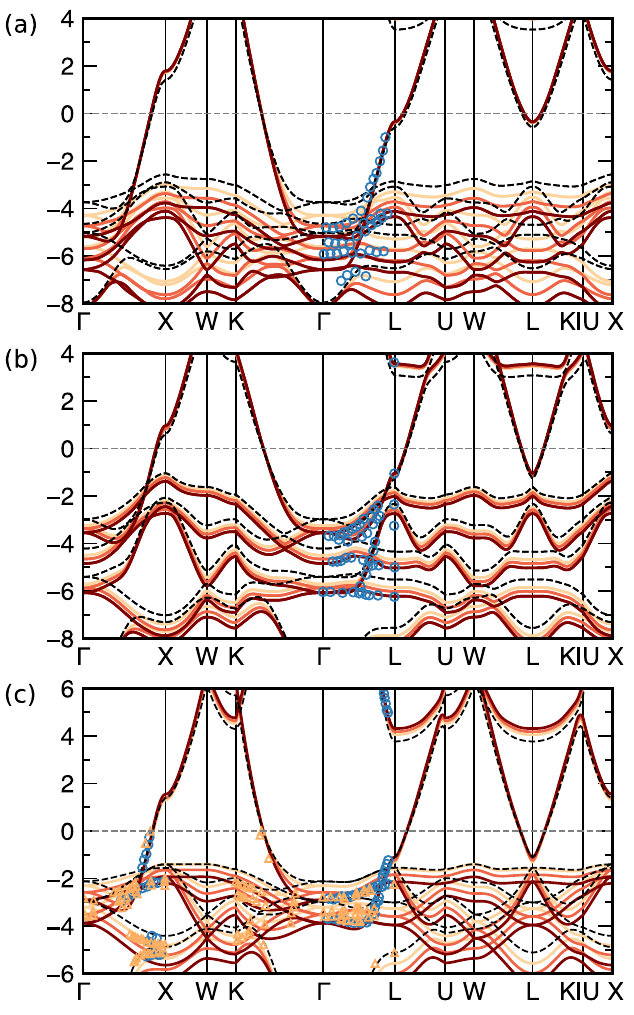}
    \caption{Calculated electronic band structures of noble metals in comparison to ARPES measurements with the same color scheme as Figure 1 in the main text. 
    The different curves are for (a) Ag, from light to dark: $U$=0,1,2 eV; (b) Au, from light to dark: $U$=0,0.5,1 eV; and (c) Cu, from light to dark: $U$=0,1,2 eV. 
    Black dashed curves are band structures calculated without quasiparticle effects and U. The agreement with experimental data \cite{wehner_valence-band_1979,mills_angle-resolved_1980,knapp_experimental_1979,courths_angle-resolved_1983} close to the top of the occupied states suggests $U$ values of 1 eV for Ag and Au, and 2 eV for Cu. 
    }
    \label{fig:bands_u}
\end{figure}

In this section, we report the electronic band structure of Ag, Au, and Cu, calculated with PBEsol+$U$+$GW$ with various $U$ to show that the $U$ parameters that we chose in the main text are reasonable. 
The band structures are shown in Figure~\ref{fig:bands_u}.
Spin-orbit interaction is included for Ag and Au. 
Overall, the choice of $U$ affects the entire set of bands for the $d$-states. The choices of $U$=1 eV for Ag and Au, and $U$=2 eV for Cu is based on agreement with ARPES data especially considering the region close to the maximum of occupied states.

\clearpage
\subsection{Divergence of second order perturbation}

\begin{figure}[!ht]
    \centering
    \includegraphics[width=0.8\columnwidth]{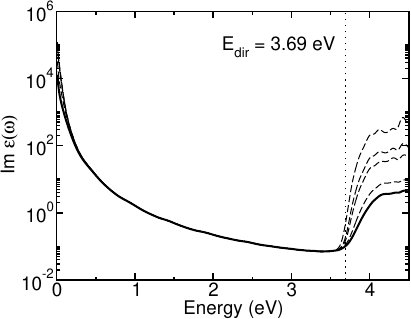}
    \caption{Calculated phonon-assisted Im $\varepsilon(\omega)$ of silver with different values for the imaginary broadening parameter $\eta$. The black solid line shows the data for $\eta$=0.1 eV, while the dashed curves from top to bottom correspond to $\eta$=0.001,0.05,0.01,0.05 eV. Strong dependence is seen at region beyond the direct onset and very close to zero (below 0.1 eV), but the spectra are independent of $\eta$ below the direct absorption onset.
    }
    \label{fig:div}
\end{figure}

In this section, we discuss our corrections to the divergence of the phonon-assisted imaginary part of dielectric function evaluated with second order perturbation theory by the inclusion of an imaginary broadening in the denominator. Due to the energy denominator for the evaluation of the generalized matrix element\cite{zhang_ab_2022}: 

\begin{align}\label{eq:indirect_S1}
\mathbf{S}_{1}\propto\sum_m\frac{\mathbf{v}_{im}(\mathbf{k})g_{mj,\nu}^\text{el-ph}(\mathbf{k},\mathbf{q})}{E_{m\mathbf{k}}-E_{i\mathbf{k}}-\hbar\omega+i\eta}
\end{align}
and
\begin{align}\label{eq:indirect_S2}
\mathbf{S}_{2}\propto\sum_m\frac{g_{im,\nu}^\text{el-ph}(\mathbf{k},\mathbf{q})\mathbf{v}_{mj}(\mathbf{k}+\mathbf{q})}{E_{m\mathbf{k}+\mathbf{q}}-E_{i\mathbf{k}}\pm\hbar\omega_{\nu\mathbf{q}}+i\eta},
\end{align}

where $E$ represent electronic energies, $\mathbf{k,q}$ represent the wave vector of the electronic and phonon, $\mathbf{v}$ represent velocity matrix elements and $g$ the excitation energy, $i,j,m$ represent initial, final, and intermediate states, and $\omega_{\nu \mathbf{q}}$ is the phonon frequency of phonon mode $\nu$. A divergence is seen when the excitation energy is very close to the energy of the intermedate state. This happens when direct absorption is allowed ($\mathbf{S}_1$) or at very low energy ($\mathbf{S}_2$). In these cases, the result depends strongly on the imaginary broadening parameter $\eta$. 
In Figure~\ref{fig:div}, we show an example with Ag, where the strong dependence is seen for the spectra below 0.1 eV and above the direct absorption onset (3.69 eV). In our study, we cut off the phonon-assisted spectra at the direct onset due to the expected dominance of the direct contribution

\clearpage

\subsection{Effects of spin-orbit coupling on band structures}
\begin{figure}[!ht]
    \centering
    \includegraphics[width=\columnwidth]{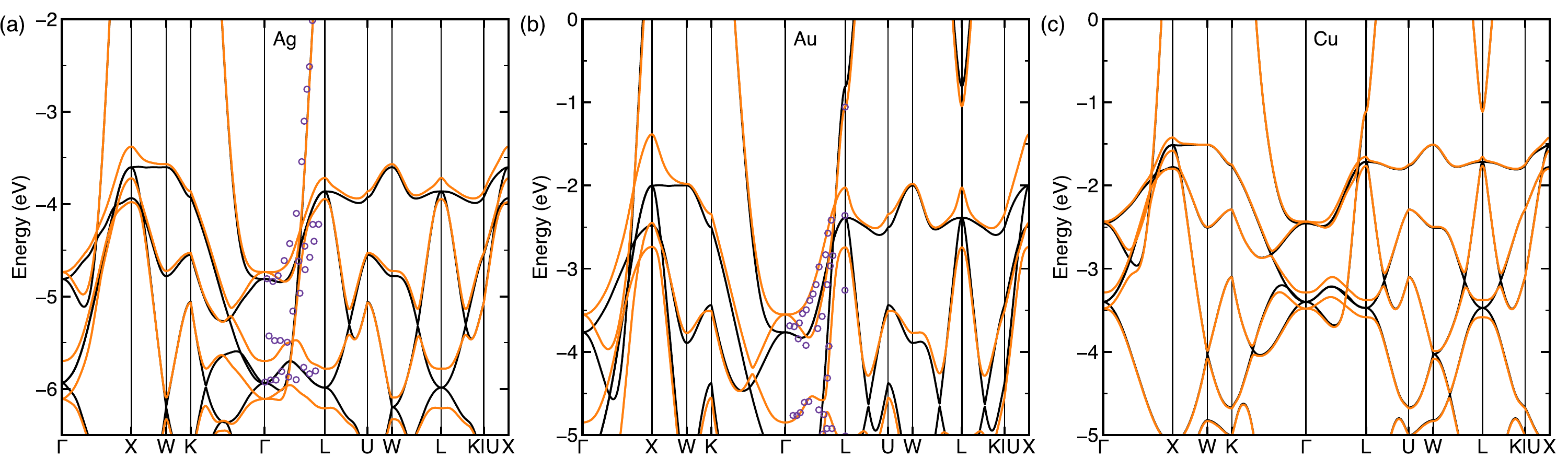}
    \caption{Electronic band structure evaluated with (orange curves) and without (black curves) spin-orbit coupling for (a) Ag and (b) Au, and (c) Cu. Significant spin-orbit splitting is seen in Ag and Au, and particularly pronounced for the latter due to the higher atomic mass.
    SOC is weak in Cu with splitting less than 0.15 eV at VBM, therefore not considered for subsequent calculations. 
    }
    \label{fig:bands_soc}
\end{figure}

In this section, we show the effects of spin-orbit coupling on the calculated electronic structure of Ag and Au. 
Figure~\ref{fig:bands_soc} shows the calculated band structure of Ag and Au with and without spin-orbit coupling. 
A significant splitting is seen in both materials. 
At the $L$-point, the splitting is 0.23 eV for Ag and 0.75 eV for Au, and at the $X$-point, the splitting is 0.34 eV for Ag and 1.07 eV for Au. 
The inclusion of SOC is also essential for the agreement between the calculation and ARPES measurements.  
We note that for Cu, a same evaluation of SOC result in only a 0.11 eV splitting at the $L$-point and 0.15 eV at the $X$-point, which is comparable to the typical value of choice for spectral broadening, therefore not included in the electronic structure and subsequent calculations due to the much higher cost of including SOC.

\clearpage

\subsection{Effects of Spin-orbit coupling on optical properties}

\begin{figure}[!ht]
    \centering
    \includegraphics[width=\columnwidth]{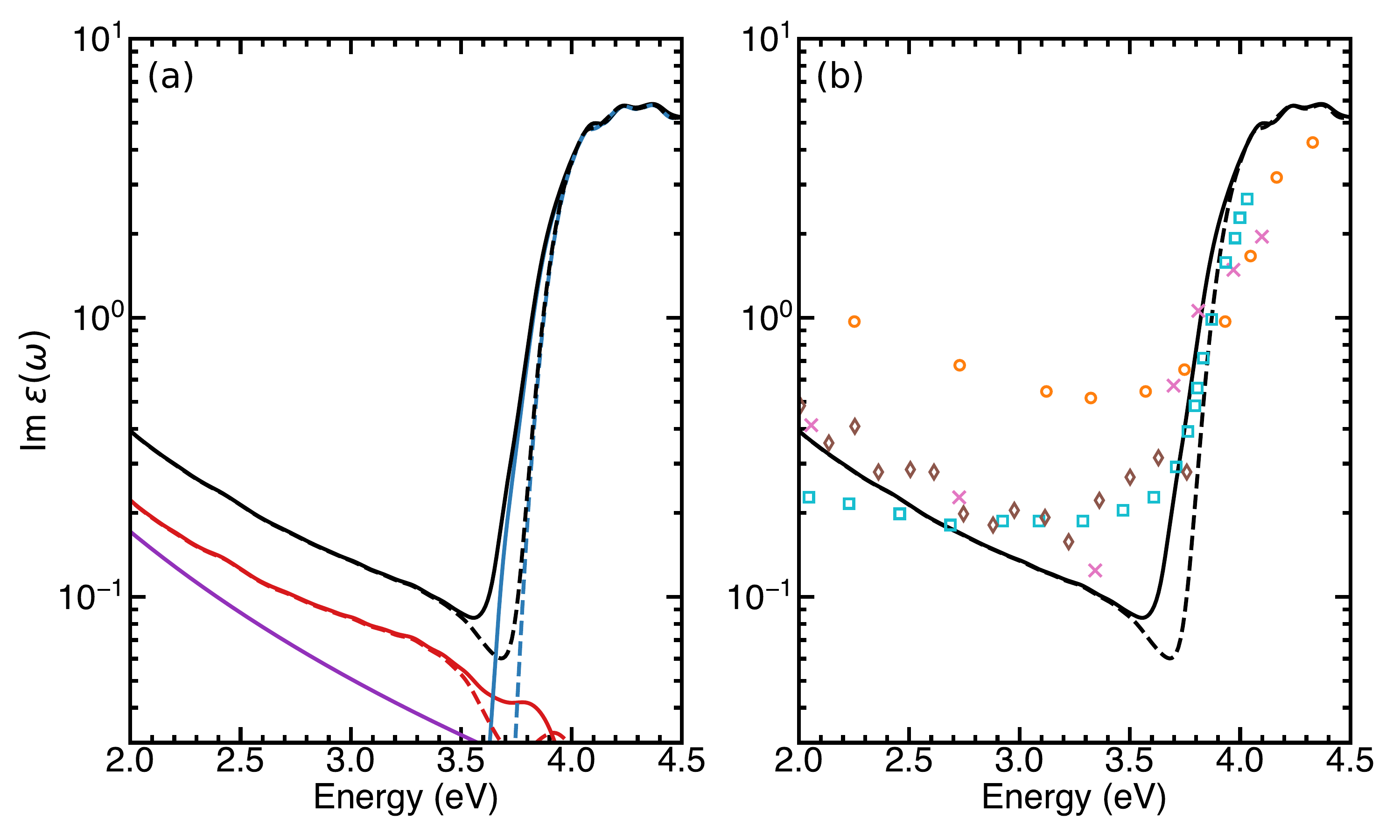}
    \caption{Effects of spin-orbit coupling on the optical spectra of Ag with (a) with SOC (solid) and without SOC (dashed) with the same color scheme as Figure 4 in the main text, breaking down by different contributions. (b) Comparison of the total curve to experimental measurements with the same color scheme as in the main text. }
    \label{fig:s4}
\end{figure}

\begin{figure}[!ht]
    \centering
    \includegraphics[width=\columnwidth]{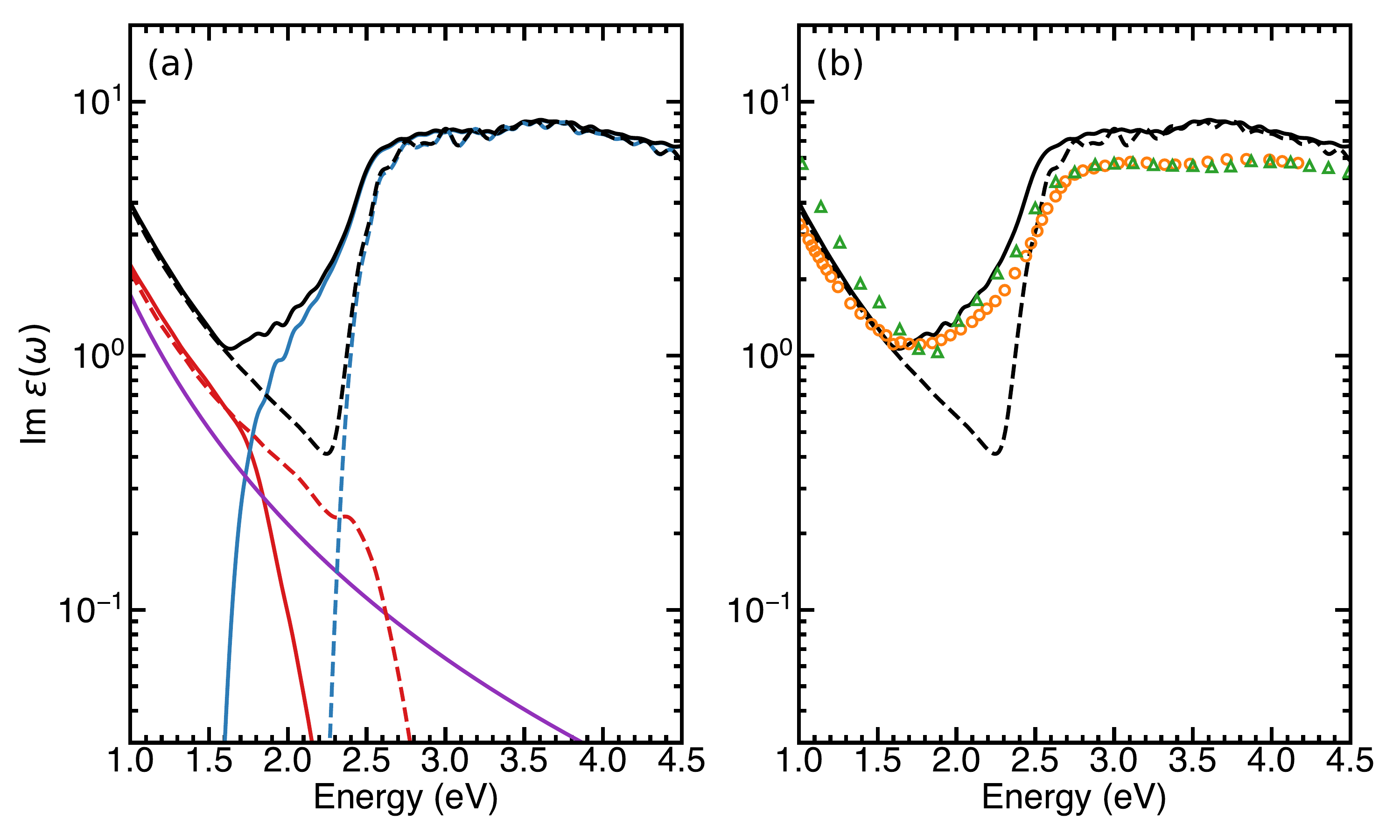}
    \caption{Effects of spin-orbit coupling on the optical spectra of Au (a) with SOC (solid) and without SOC (dashed) with the same color scheme as Figure 4, analyzed in terms of contributions by different absorption processes. (b) Comparison of the total curve to experimental measurements with the same color scheme as in the main text. SOC is essential for correct description of the absorption onset in Au. }
    \label{fig:s5}
\end{figure}

In this section, we show the effects of spin-orbit coupling on the calculated optical properties of Ag and Au. In Figure~\ref{fig:s4} and Figure~\ref{fig:s5}, we show the effects of SOC on the direct and phonon-assisted spectra for silver and gold, respectively. 
In both cases, it can be seen clearly that the effect of SOC is rather weak on the phonon-assisted contribution. 
The major effect of considering SOC is the change of the onset for the direct absorption. 
While such effects are relatively week for Ag, it becomes crucial for Au, as significantly better agreement from the experimental measurements is achieved after considering SOC, especially in the 1.5 to 2.5 eV region. 
The influence is expected for Au. 
It can be seen from the band structure that SOC induces a splitting at e.g. the $X$-point by an amount of about 2 eV, and the $L$-point by an amount of about 1 eV. 


\bibliography{metals.bib}

\end{document}